\documentclass[aps,prl,twocolumn,groupedaddress,showpacs]{revtex4}
\usepackage{graphicx}

\newcommand{\tabrule}{\rule[-0.2em]{0em}{1.2em}}
\usepackage{amsmath}
\usepackage{booktabs}
\usepackage{color}

\begin{document}

\title{Electromagnetically controlled  multiferroic thermal diode}


\author{L. Chotorlishvili, S. R. Etesami, and J. Berakdar}
\affiliation{Institut f\"ur Physik, Martin-Luther-Universit\"at Halle-Wittenberg, 06099 Halle/Saale, Germany}

\author{R. Khomeriki }
\affiliation{Physics Department, Tbilisi State University, 0128 Tbilisi, Georgia}
\author{Jie Ren$^{1,2}$}
\affiliation{$^\textit{1}$Center for Phononics and Thermal Energy Science, School of Physics Science and Engineering, Tongji University, 200092 Shanghai, China\\
$^\textit{2}$Department of Chemistry, Massachusetts Institute of Technology, 77 Massachusetts Avenue, Cambridge, MA 02139, USA}

\date{\today}

\begin{abstract}
We propose an electromagnetically tunable thermal diode based on  a two phase  multiferroics composite.
 Analytical and full numerical calculations for prototypical heterojunction composed of Iron on  Barium titanate in the tetragonal phase
  demonstrate a strong heat rectification effect that can be controlled externally
 by a moderate electric field. This finding is of an importance for thermally based information processing and sensing and
   can also be integrated in (spin)electronic circuits for heat management and recycling.
\end{abstract}

\maketitle


{\it{Introduction}}.
A diode, i.e., a device that  controls the electrical current flow direction, is an integral part of everyday electronics.
 The sonic counterpart governs the propagation of mechanical vibrations and has wide-ranging applications in acoustics,
medical sensing, and  in heat management. Acoustic (sound waves) diodes were recently   demonstrated
 \cite{N13,N14}.
Thermal diodes are more challenging, however. Even though heat and sound are of phononic nature, the frequency range of the latter
is typically in the range of kHz-GHz (hypersound). Heat, on the other hand, is mediated by a broad spectrum of THz vibrations. Controlling heat diodes is therefore more delicate, but on the plus side the relevant scale for material structuring  is on the nanometer allowing so, as shown below, to exploit the marked achievements of nanotechnology in
tuning the  material compositions and the associated electric, magnetic and optical  properties.
Applications are diverse. For instance, in spintronics it was shown that a thermal gradient may generate
a direction-dependent spin current that can be utilized  for information handling~\cite{SSE}. Such thermal magnetic diodes would add so an essential element towards thermally based spintronic circuits.
 Generally, substantial research was  devoted in recent years  to phononic-based diodes~\cite{12,13,14,15,16}.
 Our aim here is to add a new facet, namely the external control of thermal diodes via electric and/or magnetic fields. In view of an experimental implementation we consider a well-tested system composed of two-phase
 multiferroic (MF), i.e., a ferroelectric (FE) structure interfacially coupled to a ferromagnet (FM). The interfacial coupling
 renders the transmission and conversion of magnetic excitations into ferroelectric ones.
  The thermal energy in the proposed  multiferroic thermal diode is carried by elementary excitations of electric  polarization and magnetization (rather than by vibrational excitation in conventional thermal diode) both of which are susceptible to external electric or magnetic fields.  As shown below, the performance of the thermal  diode is then controllable electromagnetically.
 Multiferroics, in general,  are intensively investigated in view of a variety of applications in electronics
  and sensing \cite{1,2,3,4,5,6,7,8,9,10}. Thus, the current study augments these applications with the possibility of
  a controlled heat recycling.

\begin{center}
   \begin{figure}[h]
    \centering$
        \begin{array}{c}
    \includegraphics[width=8cm]{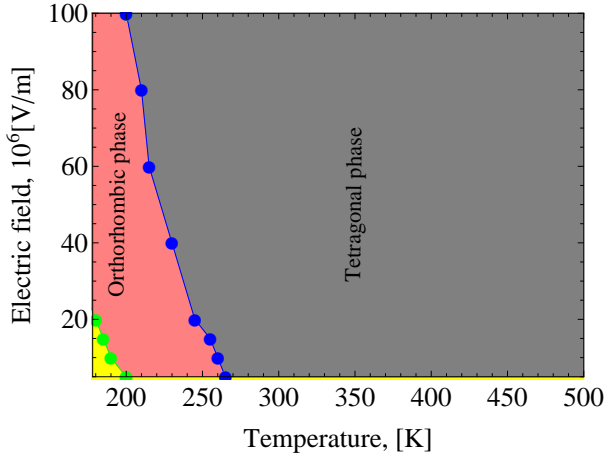}
        \end{array}$
        \caption{\label{T_E_phase_diagram}  Full-Simulations of BaTiO$_3$ phase diagram based on 8-order temperature-dependent potential\cite{Wang}.}
    \end{figure}
\end{center}

{\it{Multiferroic thermal diode}}.
The  physics of a thermal diode is a resonance phenomenon \cite{16} relying  on the overlapping of the temperature-dependent power spectra of thermal excitations (mediated by polarization, magnetization, and other type of excitations)  of the two different diode segments. In addition, the dependence of  frequency on the oscillation amplitude, i.e., the nonlinear nature of excitations is a key factor.
 A perfect thermal conductance hints on power spectra overlapping. Our aim here is to  demonstrate that thermal bias applied on the edges of  MF thermal diode  generates a heat flux that can be rectified and controlled by temperature, electric field and interface ME coupling.
To this end, we   assume that the FM part of the thermal diode is a normal ferromagnetic metal (e.g., Fe).  As a prototypical FE we employ  BaTiO$_3$.
For this experimentally realized composite  an interfacial magneto electric coupling \cite{6,17}  was demonstrated.
The ferroelectric dynamics of BaTiO$_3$ is captured by the Ginzburg-Landau-Devonshire (GLD) potential~\cite{18}
valid at temperatures $\sim$280-400 [K] (tetragonal phase) in which case the polarization switches bidirectionally.
In the spirit of a coarse-grained approach, the FE order parameter  is discretized into $N$  cells (also called sites) each with a size of 1 nm  \cite{PiLe04}. The coarse-grained polarization  at site $n$ is referred by $p_n$.
{In the tetragonal phase, realized also at room temperature, we have
one component (Ising type) polarization vector
$\vec{p}_{n}=\big(0,0,p_{n}^{z}\big),~~n=1,...N$ (here $n$ is the
site number) entering  the ferroelectric free energy functional
\cite{18}. In the context of thermal diode an important fact is that, by applying an external electric field, the temperature range of the tetragonal phase can be extended  \cite{Wang,Fesenko}. Taking the general cubic paraelectric phase as a reference, we performed  numerical calculations which turned  out to be in line with the experimentally determined phase diagram \cite{Wang}. The result of our calculations is shown in Fig.~\ref{T_E_phase_diagram}. As we see by applying an electric field with an amplitude  E=100 [MV/m] the lower limit of the tetragonal phase is reduced from T=280 [K] to the T=200 [K] while the upper limit of the tetragonal phase exceeds  T=500[K].}

For FM we employ the well-established classical Heisenberg model to describe transversal excitations
  %
of the coupled (coarse grained)  magnetic moment   $\vec{M}_n$ at
site $n$. {Experiments done for the different materials \cite{Hess}
evidence the dominant role of  magnons for the thermal heat
conductance at relatively high temperatures $T>20$ [K]. The
relevance of magnons to thermal heat conductance was also confirmed
by the spin Seebeck effect \cite{Xiao}.} The multiferroic
interaction between the interfacial FE cell (with $p_{N}$) and the
adjacent FM cell (with $M^z_{N+1}$) is described by the
$\mathcal{PT}$ invariant term
$V_{ME,m}=-g_m\left(p_{N}M^z_{N+1}\right)^m$, the form and the
origin of which we discussed at length recently and contrasted with
experimental findings \cite{jia14,19,20,21,22,23,24,11}. Here we
account for the linear ($V_{ME,1}$) and quadratic  ($V_{ME,2}$)
terms (for low energy excitations, higher order terms are less
relevant to the effects studied here). $g_m$ is the magnetoelectric
coupling constant.
The total Hamiltonian of the composite reads $H=H_{p}+H_{s}+V_{ME}$,
where
$H_{p}=\sum\limits_{n=1}^N\big[\frac{1}{2}\big(\frac{dp_{n}}{dt}\big)^{2}-\frac{\alpha^{EF}}{2}p_{n}^{2}+\frac{\beta^{EF}}{4}p_{n}^{4}+\frac{1}{2}\big(p_{n+1}-p_{n})^{2}-Ep_{n}\big]$
is the FE Hamiltonian and $H_{S}=\sum\limits_{k=N+1}^M
\big(-J\vec{M}_{k}\vec{M}_{k+1}-D\big(M_{k}^{z}\big)^{2}-BM_{k}^{z}\big)$
is the FM Hamiltonian. Unless otherwise stated, we use dimensionless
units (d.u.). For  values of all parameters in conventional units as
used experimentally   we refer to the Appendix. {The effect of the
applied thermal bias can be described by a stochatic field added to
the effective electric field in time-dependent GLD equation
\cite{Klotins}. The microscopic mechanism for the  emergence of
noise in FE is based on phonons. Thermally activated phonons lead to
electric dipole vibrations that can be captured by a random electric
field.  Experimentally, thermally activated  polarization switches
at much lower field strengths than  predicted by GLD
phenomenology\cite{Viehland} (without including noise).} The
equations of motion for the  polarization $p_n$   read~\cite{11}
\begin{equation}
\label{eq_1}
\begin{split}
 \frac{d p_n}{d t}=&q_n\\
 \frac{d q_n}{d t}=&\alpha^{FE}p_n-\beta^{FE}p^3_n-(2p_n-p_{n+1}-p_{n-1})+E\\
 &+g_1M^z_1\delta_{nN}+2g_2p_n\big(M^z_1\big)^{2}\delta_{nN}-\gamma_nq_n\delta_{1n}\\
 &+\delta_{1n}\xi_n,\quad\quad n=1,\cdots N.
  \end{split}
\end{equation}
Here $\alpha^{FE}$, $\beta^{FE}$ are the kinetic parameters of the GLD potential, $E$ is the amplitude of the external electric field, $g_1M^z_1\delta_{nN}+2g_2p_n\big(M^z_1\big)^{2}\delta_{nN}$ is the contribution from the ME coupling, and the last two terms in (1) describe the influence of the thermal bias applied on the edges of the FE chain. The correlation function of the random noise $\xi_n$ is related to the kinetic constant $\gamma_n$ and the thermal energy $k_BT$ via the Einstein relation
\begin{equation}
\label{eq_2}
    \langle\xi_m(t)\xi_m(t^\prime)\rangle=2\gamma_mT_m\delta(t-t^\prime),\quad   m=1,\cdots N.
\end{equation}

The magnetization dynamics of the FM part is governed by a set of coupled polarization-dependent LLG equations as follows
\begin{equation}
\label{eq_3}
    \frac{d \vec{M}_k}{d t}=-\frac{1}{1+\alpha_k^2}\vec{M}_k\times\left(\vec{B}^{eff}_k+\alpha_k\vec{M}_k\times\vec{B}^{eff}_k\right) .
\end{equation}
Here $\alpha_k=\alpha\delta_{kM}$ and $\vec{B}^{eff}_k$ are the total effective (electric polarization-dependent) magnetic field acting on the k-th magnetic moment $k\in [N+1,M]$
\begin{equation}
\label{eq_4}
\begin{split}
 \vec{B}^{eff}_k=&\vec{i}_zB+J\left(\vec{M}_{k-1}+\vec{M}_{k+1}\right)+\vec{i}_z2DM^z_k\\
 &+\vec{i}_zg_1p_N\delta_{k\, {N+1}}+\vec{i}_z2g_2p_N^2M^z_k\delta_{k\, {N+1}}+\delta_{kM}\vec{\eta}_k.
  \end{split}
\end{equation}
 $\vec{i}_{z}$ is a unit vector along the magnetization direction of the undistorted FM which we choose
 as the $z$ direction.
The effective magnetic field (Eq.~(\ref{eq_4})) contains a deterministic contribution from the external magnetic field $\vec{i}_zB$, and the contributions from exchange     $J\left(\vec{M}_{k-1}+\vec{M}_{k+1}\right)$ and    magnetic anisotropy $\vec{i}_z2DM^z_k$. Due to its interfacial nature the
 magnetoelectric coupling $\vec{i}_zg_1p_N\delta_{k\, {N+1}}+\vec{i}_z2g_2p_N^2M^z_k\delta_{k\, {N+1}}$ acts on the interfacial FM and FE cells
  only. The random magnetic field $\vec{\eta}_k$ enters   the dynamic of the edge cells only (thermal bias is applied at the end of the FM chain), while the heat propagation through the structure is evaluated
  self-consistently. The  random magnetic field $\vec{\eta}_k$ is quantified via the correlation function
\begin{equation}
\label{eq_5}
    \langle\eta^i_k(t)\eta^j_k(t^\prime)\rangle=2\alpha_k T_k\delta_{ij}\delta(t-t^\prime).
\end{equation}
Here $i$ and $j$ define the Cartesian components of the random magnetic field, $k$ numbers the
 cell, and $T_{k}$ is the cell-dependent local temperature.  $\alpha_k$ is the dimensionless Gilbert damping constant. Values of the FM and FE parameters used in the calculations are given in the TABLE I in the Appendix.
%
{Following the continuity equation for the local energy and the
equipartition theorem, the heat current and the temperature profile
can be evaluated self-consistently \cite{16}. In particular, the
expression for the heat current in the FE part reads
$J^H_k=-\langle\dot{p}_k(p_{k+1}-p_k)\rangle$. The time derivative
of the polarization $\dot{p}_k$ plays the role of a canonical
momentum. In the FE part the local (site dependent)
 temperature follows from its relation to the average local  kinetic energy \cite{16}
 which in our scaled units implies  $T_k=\left(\frac{dp_k}{dt}\right)^2$. We note that  average here  means
 long time average  which in
numerical simulations is implemented as  ensemble average.}
%
We derive the expression for the local heat current in the FM part by using Heisenberg equation of motion $\frac{\partial h_{k,k+1}}{\partial t}=i\left[H_S,h_{k,k+1}\right]$. Here $H_S=-J\sum_{k}\vec{M}_k\cdot\vec{M}_{k+1}-D\sum_{k}\left(M^z_k\right)^2-B\sum_{k}M^z_k$ is the Hamiltonian of the system and $h_{k,k+1}=-J\vec{M}_k\cdot\vec{M}_{k+1}-D\left(M^z_k\right)^2-BM^z_k$ is the local Hamiltonian. After straightforward calculations the heat current in the FM part is obtained as
\begin{equation}
\label{eq_8}
\begin{split}
 J^H_k=&i\left[h_{k+1,k},h_{k,k-1}\right]\\
 =&2DJ\left(M^x_{k+1}M^y_{k}M^z_{k}-M^x_{k}M^y_{k+1}M^z_{k}\right)\\
 &+DB\left(M^y_kM^x_{k+1}-M^x_kM^y_{k+1}\right)\\
 &-J^2\left(M^x_{k-1}M^y_{k}M^z_{k+1}-M^x_{k-1}M^y_{k+1}M^z_{k}\right)\\
 &-J^2\left(M^x_{k+1}M^y_{k-1}M^z_{k}-M^x_{k}M^y_{k-1}M^z_{k+1}\right)\\
 &-J^2\left(M^x_{k}M^y_{k+1}M^z_{k-1}-M^x_{k+1}M^y_{k}M^z_{k-1}\right).
  \end{split}
\end{equation}
The equilibrium temperature $T_k$ is evaluated self-consistently via the relation $ M^\|_k=L\left(\frac{\vec{M}_k\cdot\vec{B}^{eff}_k}{T_k}\right)$,
where $L(...)$ is the Langevin  function and $M^\|_k$ is the component of magnetization vector parallel to the effective field $\vec{B}^{eff}_k$.

\begin{center}
   \begin{figure}[h]
    \centering$
        \begin{array}{c}
    \includegraphics[width=\columnwidth]{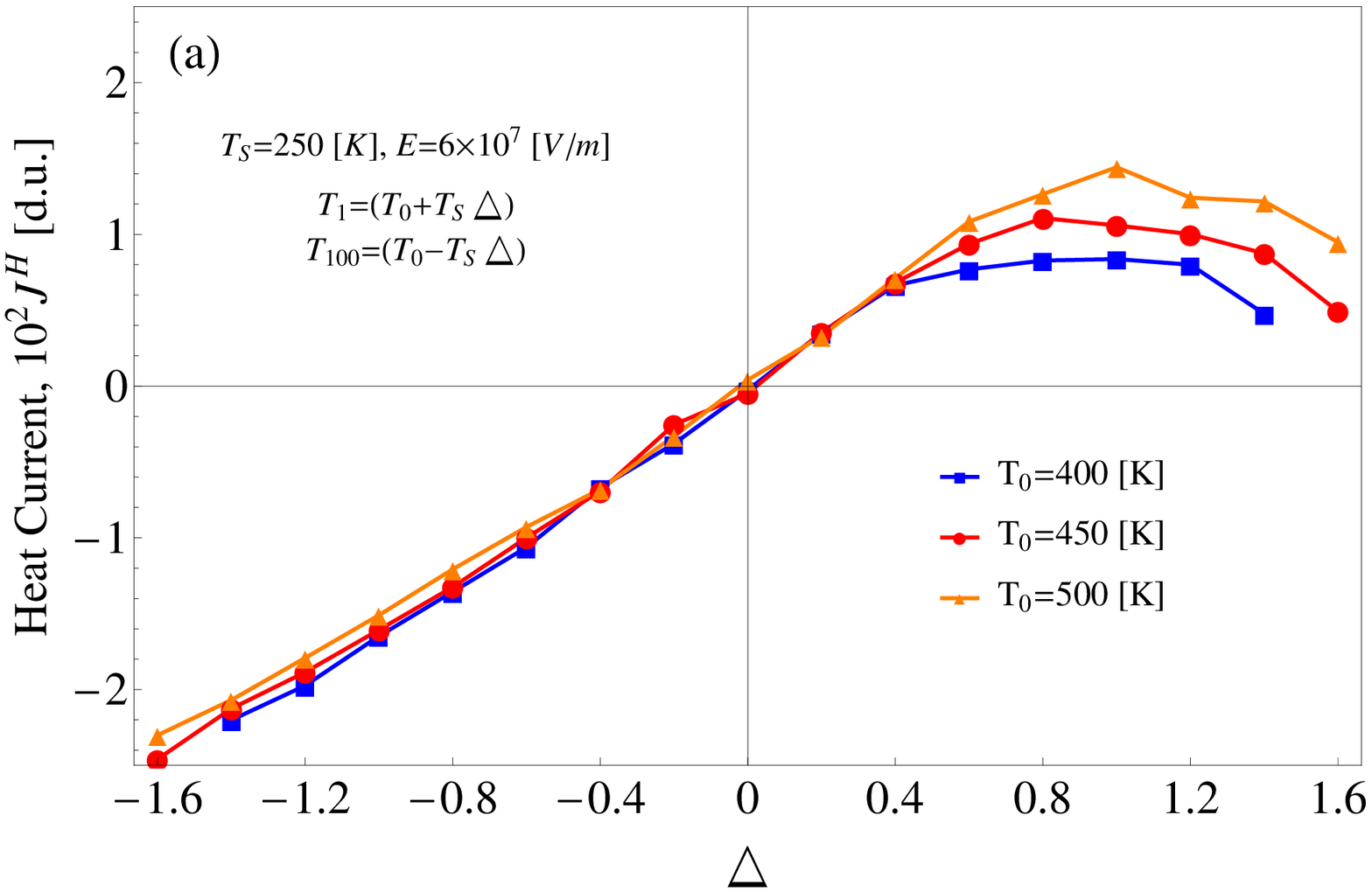}\\
    \includegraphics[width=\columnwidth]{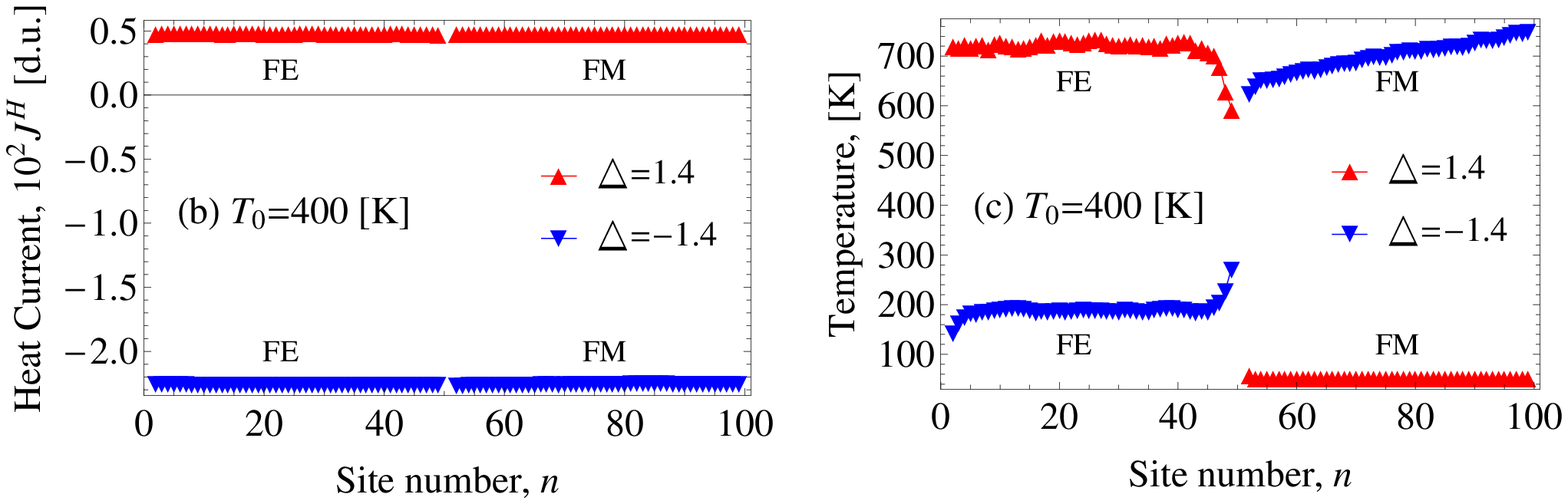}
        \end{array}$
        \caption{\label{temperature-heat_profiles} a) Diode heat current  under different edge temperatures. The multiferroic diode includes 50 FE cells (from 1 to 50) and 50 FM cells (from 51 to 100). In dimensional units, the applied electric and magnetic fields are $E=1.74$ d.u. and $B=0.0$ d.u. and FE-FM coupling coefficients are $g_1=g_2=-1$ d.u.. Other employed parameters and reduced unit coefficients are tabulated in TABLE I in the Appendix.  b) Heat current profile for forward temperature bias ($\Delta=1.4$), reverse temperature bias ($\Delta=-1.4$) and $T_0=400$ K. c) Temperature profiles for forward temperature bias ($\Delta=1.4$) and reverse temperature bias ($\Delta=-1.4$), $T_0=400$ K. In the both cases temperature formed in the FE part corresponds to the tetragonal phase.}
    \end{figure}
\end{center}
{\it{Interface effect and heat rectification}}.-
An important element of the thermal diode is the interface thermal resistance (ITR), usually referred to as the asymmetric Kapitza resistance \cite{16} as it quantifies  the asymmetry in interfacial resistance.
We will consider the cases in which the
hot thermal bath is applied to the FE part $T_{FE}>T_{FM}$ and to the FM part $T_{FE}<T_{FM}$ respectively. Inverting the sign of the thermal bias for a constant temperature difference $\Delta T =\mid T_{FE}-T_{FM} \mid$ drastically changes the heat flux $J_{+}\neq J_{-}$ and the resistance $R_{+}=\Delta T/J_{+}$, $R_{-}=\Delta T/J_{-}$. The ratio between the two different resistance $R_{+}/R_{-}$  measures  the rectification effect.
The rectification effect of the MF diode stems from  the overlapping of the spectra of the FE and FM subsystems. The frequency of the linear excitations in  FM $\omega_{FM}$ is set by the anisotropy constant $\approx 2D$ \cite{11}.
The applied electric field substantially modifies the frequency of linear excitations in the FE part, $\omega_{FE}$. Basically the electric field shifts the minimum of the GLD potential derived from the relation  $\partial_{p}H_{p}=0$. In the limit of a weak coupling between the dipoles, FE frequency takes the form $\omega_{FE}\big(E\big)=\big(4\alpha^{FE} \cos^{2}\big[\cos^{-1}\big(\frac{3|E|}{2\alpha^{FE}}\sqrt{\frac{3\beta^{FE}}{\alpha^{FE}}}\big)/3\big]-\alpha^{FE}\big)^{1/2}$. So, the  correction in the FE frequency $\Delta \omega_{FE}=\omega_{FE}\big(E\big)-\omega_{FE}\big(0\big)$ is even in the electric field (for more details we refer to the Appendix). Therefore, the heat current is symmetric with respect to the change of the electric field's sign  $E \rightarrow - E$. On the other hand, maximal heat conductance occurs when    FM and FE frequencies $\omega_{FM}\approx \omega_{FE}\big(0\big)+\Delta \omega_{FE}$ match. Thus, the electric field can be utilized to enhance the heat current.
 From the frequency matching condition and for the parameters listed in the TABLE I in the appendix, we obtain an estimation of the optimum electric field as $|E|=|\frac{2\alpha^{FE}}{3}\sqrt{\frac{\alpha^{FE}}{3\beta^{FE}}}\cos \big(\frac{3}{2}\cos^{-1}\big(\frac{4D^{2}-\alpha^{FE}}{2\alpha^{FE}}\big)\big)|=0.1$ (d.u.). In  conventional units this corresponds to an electric field of  $E=3.4\times 10^4[V/cm]$. Increasing the electric field strength results
 in a mismatch of FE and FM spectra and hence a decrease of the heat current. This analytical estimation is confirmed  by full  numerical calculations as well (Fig.~\ref{electric_filed_effect} below). Interface ME coupling leads to a small shift between analytically estimated and numerically calculated values of the optimum electric field. However, we see prominent maximum in the heat current for optimal electric field.


{\it{Temperature effects on  MF diode}}.-
We implemented full numerical simulations for a MF thermal diode consisting of 50 dipolar and 50 magnetic cells. Calculations are also done for  larger system (not shown) up to  500 dipolar and 500 magnetic cells and we did not observe significant size effects. 
Of a special interest is the rectification effect.  We present the
edge temperatures in the following form:
$T_{1}=T_0+T_{S}\Delta,~~T_{M}=T_0-T_{S}\Delta$ where $M$ is the
total number of sites. Thus, the difference between the edges
temperatures is $T_{1}-T_{M}=2 T_{S}\Delta$. Inverting  the thermal
bias  simply means $\Delta \rightarrow -\Delta$. The heat current as
a function of  $\Delta$ for  different values of $T_{0}$ is shown in
Fig.~\ref{temperature-heat_profiles}a. We observe that at larger
temperature  $T_{0}$ the asymmetry becomes stronger. The uniform
heat flux  through the system [see
Fig.~\ref{temperature-heat_profiles}b] affirms that the system is in
the nonequilibrium steady state. {On the other hand due to the
different heat capacities of the FE and FM systems and the different
heat exchange rates with the environment, the temperatures formed
self-consistently in the FE and FM parts are different.} In the case
of an applied positive thermal bias the heat flux $J_{+}=0.8$ d.u.
while for a negative thermal bias the flux reaches $J_{-}=2.5$ d.u.
and therefore  $R_{-}/R_{+}<1$. This rectification is also
characterized by distinct temperature profiles for opposite thermal
differences [see Fig.\ref{temperature-heat_profiles}c].

{\it{Electric field effect on the MF diode}}.-
The  heat current as a function of $\Delta$ is displayed in Fig.~\ref{electric_filed_effect}. We note that the
change of the sign of $\Delta$ corresponds to the inverted thermal bias. Besides, we consider different amplitudes of the applied
electric field, in order to see whether an electric field may enhance the heat rectification effect. As shown in Fig.~\ref{electric_filed_effect}a the
rectification effect becomes stronger upon  increasing the electric field strength. However, the role of the electric field is not trivial.
As  shown in the inset, there is an optimum electric filed for which the asymmetry of diode is maximal. The optimal value $E\approx 0.75$ d.u. corresponds to the frequency matching condition $\omega_{FM}\approx \omega_{FE}\big(0\big)+\Delta \omega_{FE}$ and is quite close to the analytical value that we  estimated above without interface ME coupling. Further increasing  the electric field destroys the spectra-matching condition and reduces the heat current. Magnetic field $B$ however monotonically decreases the heat flux as shown in Fig.~\ref{electric_filed_effect}b. This is due to the fact that the stronger the magnetic field the stiffer the magnetization in the FM part, which suppresses the energy transport.

\begin{center}
   \begin{figure}[t]
    \centering$
        \begin{array}{c}
    \includegraphics[width=\columnwidth]{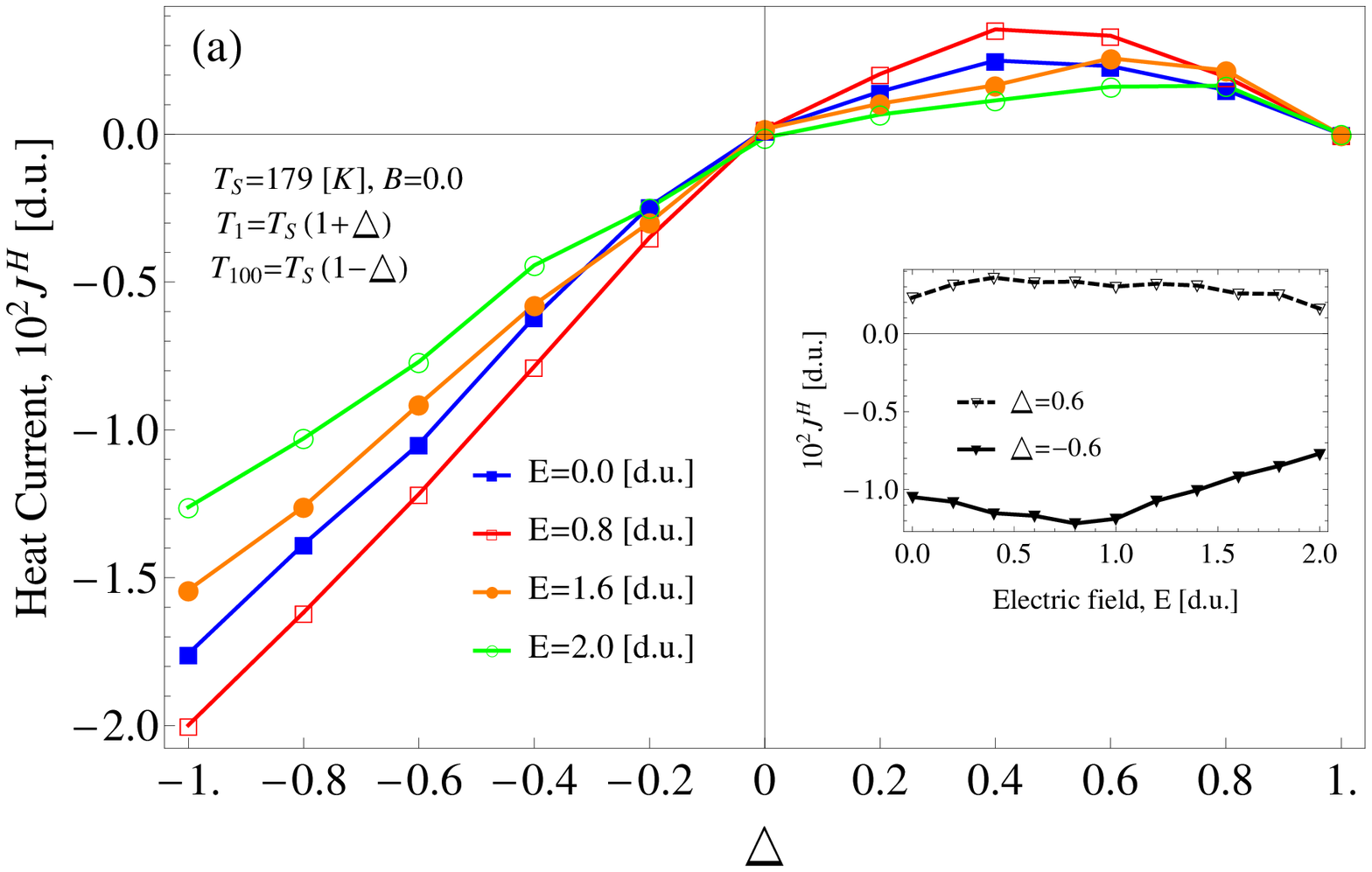}\\
    \includegraphics[width=\columnwidth]{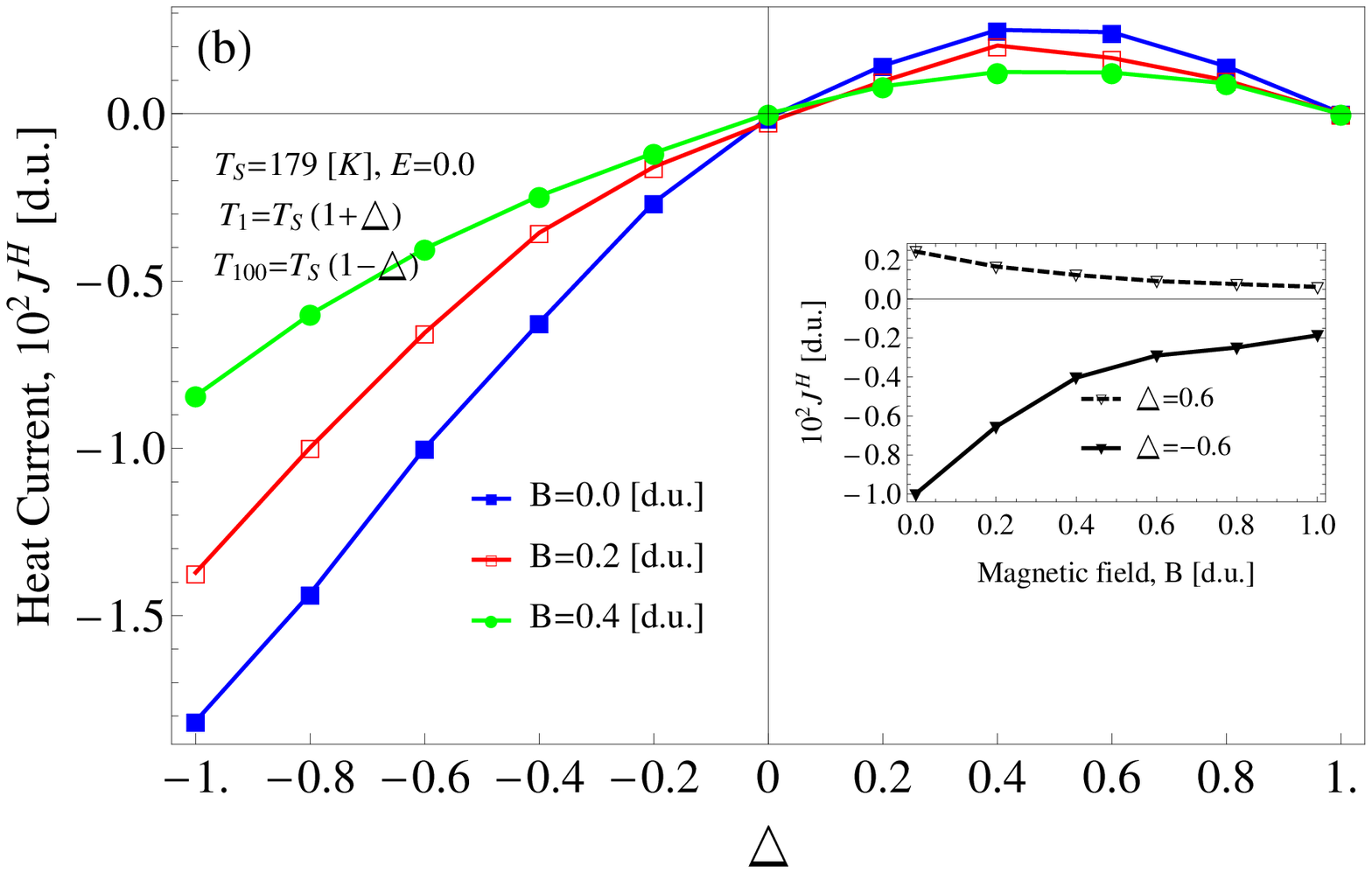}
        \end{array}$
        \caption{\label{electric_filed_effect} a) Heat current versus the biased temperature difference in the MF diode for different values of electric filed ($E$). The MF diode includes 50 FE cells (from 1 to 50) and 50 FM cells (from 51 to 100). FE-FM coupling coefficients are $g_1=g_2=-1$ d.u.. Other employed parameters are tabulated in TABLE I in the Appendix. The inset shows the dependency of heat current on electric field for $\Delta=\pm0.6$. b) The same but for magnetic field. The electric field increases the heat current
         reaching a maximum  at the optimal electric field. In contrast, the magnetic field decreases the heat current.}
    \end{figure}
\end{center}

\begin{center}
   \begin{figure}[!t]
    \centering
    \includegraphics[width=\columnwidth]{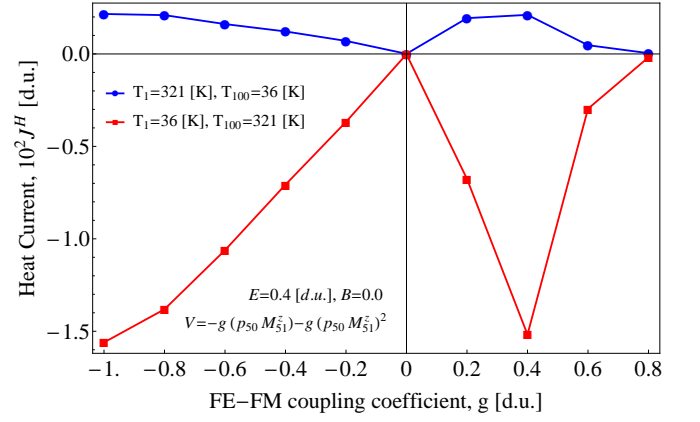}
        \caption{\label{coupling_effect} Heat current in the MF diode versus FE-FM coupling coefficients ($g=g_1=g_2$ d.u.) for forward and reverse temperature biases. The MF diode includes 50 FE cells (from 1 to 50) and 50 FM cells (from 51 to 100). The applied electric and magnetic fields are $E=0.4$ d.u. and $B=0.0$ d.u.. Other employed parameters are tabulated in TABLE I in the Appendix.}
    \end{figure}
\end{center}

{\it{Interfacial ME coupling effects on the MF diode}}.-
 Fig.~\ref{coupling_effect}  shows the heat current as a function of the interfacial ME coupling constant ($g=g_1=g_2$ d.u.) for forward and reverse biases, respectively. Without magnetoelectric coupling  the heat current across the MF diode diminishes. In the range of $g=[-1,0]$ d.u. the rectifying effect is magnified monotonically when increasing $g$ from $0$ to $-1$ and the maximum rectifying effect (asymmetry) is achieved at $g=-1$ d.u., which is the value we have used in all other simulations. Surprisingly, a different picture emerges in the positive range of $g=[0,0.8]$ d.u., where a large coupling strength near $0.8$ d.u. deteriorates the heat current in both directions. The optimal transport and rectification are achieved at an intermediate strength of $g$ around $0.4$ d.u. This distinct response upon the sign change of the coupling constant can be traced back to the fact that such a change influences the ground state of the polarization and the magnetization configurations of the MF diode system.
For a different sign of the coupling constant, slightly different configurations correspond to the ground state with a minimal energy. For the explicit form of the interface coupling term $V=-g\big(P_{N}M_{1}^{z}\big)-g\big(P_{N}M_{1}^{z}\big)^{2}$ we  find that  the positive coupling constant $g>0$    favors   large values of the magnetization component $M_{1}^{z}$. Aligning the magnetic moment along the $z$ axis naturally decreases the current, which is consistent with the picture of the inset of Fig.~\ref{electric_filed_effect}b. In the case of a negative $g<0$ the situation is different. A small $M_{1}^{z}$ means a larger transversal components and this enhances the current.
In summary, we  proposed and demonstrated a thermal diode based on a two phase multiferroic composite.
The heat transfer through it can be rectified and controlled by  a thermal bias and an electromagnetic field.
In particular, the external electric field applied to the ferroelectric part of the multiferroic thermal diode can substantially enhance the heat conductance and the rectification. On the other hand we found that an applied magnetic field decreases the heat current.
We demonstrated and  discussed how  the interfacial magnetoelectric coupling influences the thermal diode operation  in a dynamical way.
In view of contemporary advances in engineering composite multiferroic structures,  the present findings are
potentially interesting for applications, e.g., as elements in thermal switches and thermal memories~\cite{24}, and thermal management via multifunctional caloric materials~\cite{Moya}.

\section{Appendix}

\emph{{Definitions of the dimensionless units.}}
The total energy of the ferroelectric subsystem  $H_{P}$ as a function of the coarse-grained polarization $P_n$ and corresponding equations of motion read: $H_{P}=\sum_{n=1}^{N}\frac{\alpha_0}{2}\dot{P}_n^2-\frac{\alpha_1}{2}P_n^2+\frac{\alpha_2}{4}P_n^4+\frac{\kappa}{2}(P_{n+1}-P_n)^2-EP_{n}$, and $\alpha_0 \ddot{P}_n=\alpha_1 P_n-\alpha_2 P_n^3-\kappa (-(P_{n+1}-P_n)+(P_n-P_{n-1}))+E$. This equation is normalized by introducing $p_n=P_n/P_0$, $E\rightarrow E/kP_{0}$. Dividing both parts by $\kappa$  leads to the new reduced time $t'^2=t^2/(\kappa/\alpha_0)$ or $t'=\omega_0 t$, where $\omega_0=\sqrt{\frac{\kappa}{\alpha_0}}$. Finally, we obtain the equation for the polarization dynamics in fully dimensionless units for $n\neq N$ (see Eq.~(\ref{eq_1}))
with $\alpha^{FE}=\alpha_1/\kappa$, $\beta^{FE}=\alpha_2P^2_0/\kappa$.

\emph{{Time scales.}} The frequency of oscillations associated with
the mode-plasma frequency $\omega_0$ is higher than the inverse
relaxation time  $\alpha_1/\gamma_{\nu}$\cite{RiHa98} (s. TABLE
\ref{tab_1}). The overdamped case yields the Landau-Khalatnikov
equation \cite{LaKh54} employed  for modeling  the polarization
hysteresis \cite{RiHa98,PiLe04}). An overview over the real
parameters and their dimensionless counterparts for the
ferroelectric subsystem is given in TABLE \ref{tab_1}. The time
scale within the present calculations is set by the frequency
$\omega_0$ which is related to the mode-plasma frequency or the fast
oscillations (also known as "eigen-displacements" or Slater modes
\cite{21}) of the $Ti$-atom in BaTiO$_3$. \textit{Ab-initio}
calculations for BaTiO$_3$ \cite{GhCo99} yield
$\Omega_{Slater}=1519$~[cm$^{-1}$]=$286\cdot 10^{12}$~[s$^{-1}$],
the experimental values  differ slightly at $T=300$~[K] yielding
$\Omega_{Slater}=1628$~[cm$^{-1}$], as given in Ref. \cite{SeGe80}.
Finally, one can also estimate the mode-plasma frequency as
\cite{Cao94}
$\omega_0=Z^{*}_{\mathrm{Ti}}e\sqrt{\frac{1}{m_{Ti}\varepsilon_0
a^3_0}}$, where $Z^{*}_{\mathrm{Ti}}=7$, given in Ref. \cite{18} is
the Born effective charge and $m_{Ti}=47.9$~[amu]=$79,5\cdot
10^{-27}$~[kg]. For the displacement of several Angstrom
$\omega_0\approx 100\cdot 10^{12}$~[s$^{-1}$]. In our numerical
calculations the dimensionless time scales, however, with the
prefactor of $\omega_0/(2\pi)$, therefore we arrive at the
approximate value of $\sim 10^{12}$~[Hz].

\emph{{Parameters of the FM part.}} For FM part employ the
Landau-Lifshitz-Gilbert equation of motion\cite{LaLi35,Gilb55} (see
Eq. (\ref{eq_3}))
Bulk parameters for Fe: anisotropy strength $K_1\approx 5.0\cdot 10^4$~[J/m$^3$] given in Ref.~\cite{22}, the saturation magnetization $M_{S}=1.7\cdot 10^6$~[A/m] given in Ref.~\cite{22}. The Larmor (precessional) frequency in the local anisotropy field scales as $\omega_{prec}/(2\pi)=\gamma 2K_1/(M_{S})\approx 8\cdot 10^9$~[Hz], frequency associated with the relaxation scales as $\omega_{rel}/(2\pi)=\alpha_{FM}\omega_{prec}/(2\pi)\approx 0.08\cdot 10^9$~[Hz]. The autocorrelation function of the thermal fields in FE and FM part in standard units are given as $\langle\xi_k(t)\xi_k(t^\prime)\rangle=\frac{2k_{B}\gamma_{\nu}}{a_{FE}^{3}}T_k\delta(t-t^\prime)$ and $\langle\eta^i_k(t)\eta^j_k(t^\prime)\rangle=\frac{2k_{B}\alpha_{FM}}{\gamma M_{\mathrm{S}}a_{FM}^{3}}T_k\delta_{kM}\delta_{ij}\delta(t-t^\prime)$ where $T_{k}$ is the site-dependent local temperature in Kelvin.\\
\emph{{Shift of ferroelectric frequency.}} We consider one unit cell
in the FE Hamiltonian: $\displaystyle
H_{P}=\frac{1}{2}\dot{p}^2-\frac{\alpha^{FE}}{2}p^2+\frac{\beta^{FE}}{4}p^4-Ep$
Equilibrium properties are given by the condition $\partial H_{p}=0$. After solving cubic equation we obtain : $p_1^{(0)}=\frac{2}{\sqrt{3}}\sqrt{\frac{\alpha^{FE}}{\beta^{FE}}}\cos\left(\frac{\theta}{3}\right)$, $p_2^{(0)}=\frac{2}{\sqrt{3}}\sqrt{\frac{\alpha^{FE}}{\beta^{FE}}}\cos\left(\frac{\theta}{3}+\frac{2\pi}{3}\right)$ and $ p_3^{(0)}=\frac{2}{\sqrt{3}}\sqrt{\frac{\alpha^{FE}}{\beta^{FE}}}\cos\left(\frac{\theta}{3}+\frac{4\pi}{3}\right)$. Here $\theta=\arccos\left(\frac{3E}{2\alpha^{FE}}\sqrt{\frac{3\beta^{FE}}{\alpha^{FE}}}\right)$ and minimum of the energy reads: $H\left(p_{1,2}^{(0)}(E)\right)=-\frac{(\alpha^{FE})^2}{4\beta^{FE}}\pm\sqrt{\frac{\alpha^{FE}}{\beta^{FE}}}E$. As we see if $E>0$ then minimum of the energy corresponds to the solution $p_1^{(0)}(E)$ while if $E<0$ then energy minimum corresponds to the solution $p_2^{(0)}(E)$.
Taking into account fact that system is even in electric field we express minimum of the energy in the form valid for the both $E>0$ and $E<0$ cases: $H\left(p_{1,2}^{(0)}(E)\right)=-\frac{(\alpha^{FE})^2}{4\beta^{FE}}-\sqrt{\frac{\alpha^{FE}}{\beta^{FE}}}|E|$.

\begin{table}[t]
\caption{Parameters of an unstrained bulk BaTiO$_3$-single crystal\cite{HlMa06,MaHl08,SeDa06,HlPe06} and  bulk bcc-Fe\cite{22} (p. 385).}
\centering
\vspace{2.ex}
\begin{tabular}{c|c|c}
\hline
\hline
parameter \tabrule & SI units & dimensional unit (d.u.) \\ \hline
$P_{0}$ \tabrule & 0.265  [C/m$^2$] & $p_n=P_n/P_0$\\
$\alpha_1$ \tabrule & $2.770\cdot 10^7$ [V$\cdot$m/C] & $\alpha^{FE}=\frac{\alpha_1}{\kappa}\approx 0.213$\\
$\alpha_2$ \tabrule & $1.7\cdot 10^8$ [V$\cdot$m$^5$/C$^3$]  & $\beta^{FE}=\frac{\alpha_2P_0^2}{\kappa}\approx 0.0918$ \\
$\gamma_{\nu}$ \tabrule & $2.5\cdot 10^{-5}$ [V$\cdot$m$\cdot$s/C]  & $\gamma_m=\frac{\gamma_{\nu}\omega_0}{\kappa}\approx 0.192$ \\
$a_{FE}$  \tabrule & $1.02\cdot 10^{-9}$ [m]& - \\
$\kappa$\tabrule &  $1.3\cdot 10^8$  [V$\cdot$m/C] & 1  \\
$E$ \tabrule &  parameter [V/m] & $E\rightarrow \frac{1}{\kappa P_0}E\approx3.4\times10^7E$  \\
$T$  \tabrule &  parameter [K] & $T\rightarrow \frac{k_B}{\kappa P_0^2a_{FE}^3}T\approx1.4\times10^{-3}T$  \\
$J$\tabrule &  -  [Joule/s]  & $J\rightarrow \frac{1}{\kappa P_0^2\omega_0a_{FE}^3}J\approx10^{8}J$  \\ \hline \hline
$M_{\mathrm{S}}$ \tabrule & $1.71\cdot 10^6$~[A/m]  & $\vec{s}_k=\vec{M}_k/M_{S}=\left(\vec{S}_k/S\right)$\\
$\gamma$ \tabrule & $ 1.76\cdot 10^{11}$  [(T$\cdot$s)$^{-1}$] & - \\
$a_{FM}$  \tabrule & $ 1.0\cdot 10^{-9}$ [m] & - \\
$\mu_{S}=M_{S}a^3_{\mathrm{FM}}$  \tabrule & $1.71\cdot 10^{-21}$ [J/T]& - \\
$\alpha_{FM}$ \tabrule & $1.0$ & - \\ 
$K_1$ \tabrule & $2.0 \cdot 10^{6}$ [J/m$^3$] & $D=\frac{\gamma a_{FM}^3K_1}{\omega_0 \mu_{S}}=0.206$ \\
$A$  \tabrule &  $2.1 \cdot 10^{-11}$ [J/m]&  $J=\frac{\gamma a_{FM} A}{\omega_0 \mu_{S}}=2.16$ \\
$B$  \tabrule &  parameter [T] & $B\rightarrow \frac{\gamma}{\omega_0}B\approx0.17B$  \\
$T$  \tabrule &  parameter [K]& $T\rightarrow \frac{k_B \gamma}{\omega_0 \mu_{S}}T\approx1.4\times10^{-3}T$  \\
$J$  \tabrule &  - [Joule/s]& $J\rightarrow \frac{\gamma}{\omega_0^2 \mu_{S}}J\approx10^{8}J$  \\ \hline \hline
\end{tabular}
\label{tab_1}
\end{table}
In order to evaluate dependence of the FE frequency on the applied external electric field we expand Hamiltonian $H_{p}$ in the vicinity of the equilibrium points.
In the equation of motion governed by linearized Hamiltonian $\ddot{p}=-\left(-\alpha^{FE}+3\beta^{FE}p_0^2\right)p+E$ enters electric field dependent frequency: $\omega^2_p(E)=\left(-\alpha^{FE}+3\beta^{FE}p_0^2\right)$. Considering small electric field $E$  in the  first order approximation from $p_{1,2,3}^{(0)}$ we obtain: $\omega_p(E>0)=\sqrt{2\alpha^{FE}}+\frac{3E}{2\alpha^{FE}}\sqrt{\frac{\beta^{FE}}{2}}$, $\omega_p(E<0)=\sqrt{2\alpha^{FE}}-\frac{3E}{2\alpha^{FE}}\sqrt{\frac{\beta^{FE}}{2}}$ and $\omega_p(E=0)=\sqrt{2\alpha^{FE}}$. FE frequency shift due to the applied weak electric field reads : $\Delta\omega_p(E)=\omega_p(E)-\omega_p(0)\approx\frac{3|E|}{2\alpha^{FE}}\sqrt{\frac{\beta^{FE}}{2}}$. As we see frequency shift is even in electric field. On the other hand the FM frequency is equal to $\omega_D=2D$. Matching condition between the frequencies ($\omega_D=\omega_p+\Delta\omega_p$) defines optimum electric field relevant to the maximal conductance: $|E|=\bigg|\frac{2\alpha^{FE}}{3}\sqrt{\frac{\alpha^{FE}}{3\beta^{FE}}}\cos\bigg(\frac{3}{2}\cos^{-1}\bigg(\frac{4D^{2}-\alpha^{FE}}{2\alpha^{FE}}\bigg)\bigg)\bigg|.$


\begin{thebibliography}{99}

\bibitem{N13}  B. Liang, B. Yuan, and J. C.  Cheng, Phys. Rev. Lett. {103}, 104301 (2009).
\bibitem{N14}
B. Liang, X. Guo, J. Tu, D. Zhang, and J. C. Cheng, Nature Mater.
{9}, 989 (2010); B. Li, {9}, 962–963 (2010); X.-F. Li, X. Ni, L.
Feng, M.-H. Lu, C. He, and Y.-F. Chen, Phys. Rev. Lett. {106} 084301
(2011); M. Maldovan, Nature {503}, 209 (2013).
%
\bibitem{SSE} J. Ren, Phys. Rev. B {88}, 220406(R) (2013); J. Ren and J.-X. Zhu, Phys. Rev. B {88}, 094427 (2013); J. Ren, J. Fransson, and J.-X. Zhu, Phys. Rev. B {89}, 214407 (2014)
%
\bibitem{12}
S. Lepri, R. Livi, and A. Politi, Phys. Rev. Lett. {78}, 1896
(1997); T. S. Komatsu, and N. Ito,  Phys. Rev. E {83}, 012104
(2011); T. S. Komatsu and N. Nakagawa, Phys. Rev. E {73}, 065107(R)
(2006).
%
\bibitem{13}
P. Kim,  L. Shi, A. Majumdar, and P. L. Mc Euen, Phys. Rev. Lett.
{87}, 215502 (2001); W.  Kobayashi, Y. Teraoka, and I. Terasaki,
Appl. Phys. Lett. {95}, 171905 (2009); B. Li, J. H. Lan, and L.
Wang, ,  Phys. Rev. Lett. {95}, 104302 (2005).
%
\bibitem{14}
M. Terraneo, M. Peyrard, and G. Casati, Phys. Rev. Lett. {88},
094302 (2002); G. Casati, Chaos {15}, 015120 (2005).
%
\bibitem{15}
B. Li, L. Wang, and G. Casati, Phys. Rev. Lett. {93}, 184301 (2004);
B. Hu, L. Yang, and Y. Zhang, Phys. Rev. Lett. {97}, 124302 (2006);
%
\bibitem{16}
N. Li, J. Ren, L. Wang, G. Zhang, P. H\"anggi, B. Li, Rev. Mod.
Phys. {84}, 1045 (2012).
%
\bibitem{1}
W. Eerenstein, N. D. Mathur, and J. F. Scott, Nature (London) {442},
759 (2006); Y. Tokura and S. Seki, Adv. Mater. {22}, 1554 (2010); C.
A. F. Vaz, J. Hoffman, Ch. H. Ahn, and R. Ramesh, Adv. Mater. {22},
2900 (2010); F. Zavaliche, T. Zhao, H. Zheng, F. Straub, M. P. Cruz,
P.-L.Yang, D. Hao, and R. Ramesh, Nano Lett.{7}, 1586 (2007).
%
\bibitem{2}
H. L. Meyerheim, F. Klimenta, A. Ernst, K. Mohseni, S. Ostanin, M.
Fechner, S. Parihar, I.V.Maznichenko, I. Mertig, and J. Kirschner,
Phys. Rev. Lett. {106}, 087203 (2011); R. Ramesh and N. A. Spaldin,
Nat. Mater. {6}, 21 (2007).
%
\bibitem{3}
M. Bibes and A. Barthelemy, Nat. Mater. {7}, 425 (2008); M. Gajek,
M. Bibe, S. Fusil, K. Bouzehouane, J. Fontcuberta, A. Barthelemy,
and A. Fert, Nat. Mater. {6}, 296 (2007).
%
\bibitem{4}
D. Pantel, S. Goetze, D. Hesse, and M. Alexe, Nat. Mater. {11}, 289
(2012); C.-W. Nan, M. I. Bichurin, S. Dong, D. Viehland, and G.
Srinivasan, J. Appl. Phys. {103}, 031101 (2008).
%
\bibitem{5}
N. Spaldin and M. Fiebig, Science {309}, 391 (2005); M. Fiebig, J.
Phys. D {38}, R123 (2005).
%
\bibitem{6}
C.-G. Duan, S. S. Jaswal, and E.Y. Tsymbal, Phys. Rev. Lett. {97},
047201 (2006).
%
\bibitem{7}
M. Fechner, I. V. Maznichenko, S. Ostanin, A. Ernst, J. Henk, and I.
Mertig, Phys. Status Solidi B {247}, 1600 (2010).
%
\bibitem{8}
M. Mostovoy, Phys. Rev. Lett. {96}, 067601 (2006); H. Katsura, N.
Nagaosa and A. V. Balatsky,  Phys. Rev. Lett. {95}, 057205 (2005);
S. Park, Y. J. Choi, C. L. Zhang, and S-W. Cheong, Phys. Rev. Lett.
{98}, 057601 (2007).
%
\bibitem{9}
M. Azimi, L. Chotorlishvili, S. K. Mishra, T. Vekua, W. Hübner and
J. Berakdar, New J. of Physics {16}, 063018 (2014).
%
\bibitem{10}
M. Azimi, L. Chotorlishvili, S. K. Mishra,S. Greschner, T. Vekua,
and J. Berakdar, Phys. Rev. B {89}, 024424 (2014)
%
\bibitem{17}
S. Sahoo, S. Polisetty, C.-G.Duan, S. S. Jaswal, E. Y. Tsymbal, and
C. Binek, Phys. Rev. B {76}, 092108 (2007).
%
\bibitem{18}
Physics of Ferroelectrics, edited by K. Rabe, Ch. H. Ahn, and J.-M. Triscone (Springer, Berlin 2007).
%
\bibitem{PiLe04}
A. Picinin, M. H. Lente, J. A. Eiras, and J. P. Rino, Phys. Rev. B
{69}, 064117 (2004)
%
\bibitem{Wang}
J. J. Wang, P. P. Wu, X. Q. Ma, and L. Q. Chen1, J. Appl. Phys.
{108}, 114105 (2010).

\bibitem{Fesenko}
O. E. Fesenko and V. S. Popov, Ferroelectrics {37}, 729 (1981).

\bibitem{Hess}
C. Hess, P. Ribeiro, B. Büchner, H. ElHaes, G. Roth, U. Ammerahl,
and A. Revcolevschi Phys. Rev. B {73}, 104407 ( 2006);  C. Hess, H.
ElHaes, A. Waske, B. Büchner, C. Sekar, G. Krabbes, F.
Heidrich-Meisner, and W. Brenig Phys. Rev. Lett. {98}, 027201
(2007).
%
\bibitem{Xiao}
J. Xiao, G. E. W. Bauer, K.-c. Uchida, E. Saitoh, and S. Maekawa,
Phys. Rev. B {81}, 214418 (2010); K.I. Uchida, T. Kikkawa, A. Miura,
J. Shiomi, and E. Saitoh, Phys. Rev. X {4}, 041023 (2014).
%
\bibitem{jia14} C.-L. Jia, T.-L. Wei, C.-J. Jiang, D.-S. Xue, A. Sukhov, and J. Berakdar, Phys. Rev. B, {\bf 90}, 054423 (2014);  A. Sukhov \emph{et al.}  Phys. Rev. B {90}, 224428 (2014);  Jedrecy, N. \emph{et al.} Phys. Rev. B {88}, 121409 (2013); T. Nan,  \emph{et al.} 
     { Sci. Rep.} {4}, 3688 (2014); C.-L. Jia, F. Wang, C.-J. Jiang, J. Berakdar, D.-S. Xue  { Sci. Rep.} {5}, 11111 (2015)
\bibitem{19}
L. Chotorlishvili, Z. Toklikishvili, V. K. Dugaev, J. Barnas, S.
Trimper, and J. Berakdar, Phys. Rev. B {88}, 144429 (2013).
%
\bibitem{20}
S. R. Etesami, L. Chotorlishvili, A. Sukhov, and J. Berakdar, Phys.
Rev. B {90}, 014410 (2014).
%
\bibitem{21}
J. Hlinka, P. Marton, Phys. Rev. B {74}, 104104 (2006); P. Marton,
J. Hlinka, Ferroelectrics {373}, 139 (2008); N. Setter, D.
Damjanovic, L. Eng, G. Fox, S. Gevorgian, S. Hong, A. Kingon, H.
Kohlstedt, N.Y. Park, G.B. Stephenson, I. Stolichnov, A.K.
Tagantsev, D.V: Taylor, T. Yamada, J. Appl. Phys. {100}, 051606
(2006); S. Sivasubramanian, A. Widom, Y.N. Srivastava,
Ferroelectrics {300}, 43 (2004).
%
\bibitem{22}
J. M. D. Coey, Magnetism and Magnetic Materials, Cambridge University Press, Cambridge (2010).
%
\bibitem{23}
A.Kl\"umper and K. Sakai, J. Phys. A Math. Gen. {35}, 2173 (2002).
%
\bibitem{24}
S. Li, X. Ding, J. Ren, X. Moya, J. Li, J. Sun, and E. K. H. Salje,
Sci. Rep. {4}, 6375 (2014).
%
\bibitem{11}
L. Chotorlishvili, R. Khomeriki, A. Sukhov, S. Ruffo, and J.
Berakdar, Phys. Rev. Lett. {111}, 117202 (2013)
%
%
\bibitem{Klotins}
S. Nambu and D. A. Sagala, Phys. Rev. B {50}, 5838 (1994); S
Sivasubramanian, A. Widom, Y.N. Srivastava  Ferroelectrics, {300},
43 (2004); E. Klotins, Ferroelectrics, {370}, 184 (2008).
%
\bibitem{Viehland}
D. Viehland, Y. H. Chen J. Appl. Phys. {88}, 6696 (2000).
%
\bibitem{Moya} X. Moya, S. Kar-Narayan and N. D. Mathur, Nat. Mater. {\bf13}, 439 (2014).
%
\bibitem{LeCa10}
J. Lee, N. Sai, T. Cai, Q. Niu, A. A. Demkov, Phys. Rev. B {\bf 81}, 144425 (2010).
%
\bibitem{RiHa98}
D. Ricinschi, C. Harnagea, C. Papusoi, L. Mitoseriu, V. Tura, M. Okuyama, J. Phys.: Condens. Matter {\bf 10}, 477 (1998).

\bibitem{LaKh54}
L.D. Landau, I.M. Khalatnikov, Dokl. Akad. Nauk SSSR {\bf 46}, 469 (1954).

\bibitem{GhCo99}
Ph. Ghosez, E. Cockayne, U.V. Waghmare, K.M. Rabe, Phys. Rev. B {\bf 60}, 836 (1999).

\bibitem{SeGe80}
J.L. Servoin, F. Gervais, A.M. Quittet, Y. Luspin, Phys. Rev. B {\bf 21}, 2038 (1980).

\bibitem{Cao94}
W. Cao, J. Phys. Soc. Jap. {\bf 63}, 1156 (1994).

\bibitem{LaLi35}
L.D. Landau, E.M. Lifshitz, Phys. Z. Sowjetunion {\bf 8}, 153 (1935).

\bibitem{Gilb55}
T.L. Gilbert, Phys. Rev. {\bf 100}, 1243 (1955) (abstract only); IEEE Trans. Magn. {\bf 40}, 3443 (2004).

\bibitem{DoFi97}
J.L. Domann, D. Fiorani, E. Tronc, Adv. Chem. Phys. {\bf 98}, 283 (1997).

\bibitem{HlPe06}
J. Hlinka, J. Petzelt, S. Kamba, D. Noujni, T. Ostapchuk, Phase Transitions {\bf 79}, 41 (2006).

\bibitem{HlMa06}
J. Hlinka, P. Marton, Phys. Rev. B {\bf 74}, 104104 (2006).

\bibitem{MaHl08}
P. Marton, J. Hlinka, Ferroelectrics {\bf 373}, 139 (2008).

\bibitem{SeDa06}
N. Setter, D. Damjanovic, L. Eng, G. Fox, S. Gevorgian, S. Hong, A. Kingon, H. Kohlstedt, N.Y. Park, G.B. Stephenson, I. Stolichnov, A.K. Tagantsev, D.V: Taylor, T. Yamada, J. Appl. Phys. {\bf 100}, 051606 (2006).



\end{thebibliography}
\end{document}